\documentclass[aps,prb,preprint,groupedaddress]{revtex4}

\usepackage{graphicx}
\usepackage{dcolumn}
\usepackage{bm}
\usepackage{amsmath,amssymb}

\def\vS{\mbox{\boldmath $S$}}

\begin{document}


\title{Magnetization Plateau of the Distorted Diamond Spin Chain}



\author{Yuki \textsc{Ueno}$^{1}$, Tomosuke \textsc{Zenda}$^{1}$, 
Yuta \textsc{Tachibana}$^1$, Kiyomi \textsc{Okamoto}$^1$ and T\^oru \textsc{Sakai}$^{1,2}$}
\affiliation{
$^{1}$Graduate School of Material Science, University of Hyogo, Hyogo 678-1297, Japan \\
$^{2}$National Institutes for Quantum and Radiological Science and Technology (QST), SPring-8, Hyogo 679-5148, Japan
}


\date{Received September 1, 2019}

\begin{abstract}

The frustrated quantum spin system on the distorted diamond chain lattice 
is investigated using the numerical diagonalization of finite-size clusters 
and the level spectroscopy analysis. 
In the previous work this system was revealed to exhibit the 1/3 
magnetization plateau due to two different mechanisms depending on 
the coupling parameters, and the phase diagram at the 1/3 magnetization 
was obtained. 
In the present work it is found that the 1/3 magnetization plateau 
vanishes for sufficiently large $XY$-like coupling anisotropy. 
The phase diagram based on the level spectroscopy analysis 
is also presented.
\end{abstract}


\maketitle


\section{Introduction}

The magnetization plateau is one of interesting phenomena in the field 
of the strongly correlated electron systems. 
It possibly appears as the quantization of magnetization, 
when the one-dimensional quantum spin system satisfies the 
necessary condition
\begin{eqnarray}
S-m={\rm integer},
\label{condition}
\end{eqnarray}
where $S$ is the total spin and $m$ is the magnetization per 
unit cell\cite{oshikawa}. 
The S=1/2 distorted diamond spin chain\cite{okamoto0} is a strongly frustrated 
quantum spin system which exhibits the 1/3 magnetization plateau 
according to the condition. 
This system was proposed as a good theoretical model of 
the compound Cu$_3$(CO$_3$)$_2$(OH)$_2$, called azurite\cite{kikuchi}.
Actually the magnetization measurement of the azurite 
detected a clear magnetization 
plateau at 1/3 of the saturation magnetization. 
The theoretical study by the numerical exact diagonalization\cite{okamoto1} 
suggested that the 1/3 magnetization plateau is induced by two different mechanisms. One is based on the ferromagnetic mechanism and the other is due to the formation of singlet dimers and free spins. 
In the case of the azurite the 1/3 plateau is supposed to 
be due to the latter mechanism. 
In some previous theoretical works\cite{okamoto2,sakai1,sakai2,ito} 
on quantum spin systems 
it was reported that the magnetization plateau disappears in the presence 
of sufficiently large $XY$-like (easy-plane) coupling anisotropy. 
Thus in this paper we introduce the $XY$-like anisotropy to the 
$S=1/2$ distorted diamond spin chain and consider the stability 
of the 1/3 magnetization plateau against the anisotropy 
in the case of both mechanisms. 
We note that the anisotropy inversion phenomena was found in the distorted diamond chain
with the $XXZ$ anisotropy\cite{oka-ichi,tokuno}.
For this purpose we use the numerical exact diagonalization of finite-size 
clusters and the level spectroscopy analysis to find the quantum phase 
transition between the plateau and the no-plateau phases.

\section{Model}

\begin{figure}[tbh]
\centerline{\includegraphics[scale=0.5]{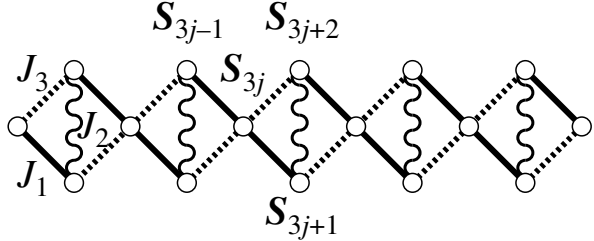}}
\caption{The model of the $S=1/2$ distorted diamond spin chain. 
}
\label{model}
\end{figure}

We investigate the model described by the Hamiltonian
\begin{eqnarray}
  &&{\cal H} = {\cal H}_0 + {\cal H}_{\rm Z} \\
  &&{\cal H}_0
   =  J_1 \sum_{j=1}^{N/3} \left[ (\vS_{3j-1} \cdot \vS_{3j})_\lambda 
          + (\vS_{3j} \cdot \vS_{3j+1})_\lambda \right]
    + J_2 \sum_{j=1}^{N/3} (\vS_{3j+1} \cdot \vS_{3j+2})_\lambda \nonumber \\
  &&~~~~~~~~~~~ + J_3 \sum_{j=1}^{N/3} \left[ (\vS_{3j-2} \cdot 
  \vS_{3j})_\lambda
          + (\vS_{3j} \cdot \vS_{3j+2})_\lambda \right] \\
  &&{\cal H}_{\rm Z}
   = -H \sum_{l=1}^{N} S_l^z \\
  &&(\vS_l \cdot \vS_m)_\lambda \equiv S_l^x S_m^x + S_l^y S_m^y + \lambda S_l^z S_m^z 
   \label{eq:ham}
\end{eqnarray}
where $\vS_j$ is the spin-1/2 operator, $J_1$, $J_2$, $J_3$ are the 
coupling constants of the exchange interactions, and $\lambda$ is the 
coupling anisotropy. 
The schematic picture of the model is shown in Fig. \ref{model}. 
In this paper we consider the case of the $XY$-like (easy-plane) 
anisotropy, namely $\lambda <1$. 

$N$ is the number of spins and $L$ is defined as the number 
of the unit cells, namely $N=3L$. 
For $L$-unit systems, 
the lowest energy of ${\cal H}_0$ in the subspace where 
$\sum _j S_j^z=M$, is denoted as $E(L,M)$. 
The reduced magnetization $m$ is defined as $m=M/M_{\rm s}$, 
where $M_{\rm s}$ denotes the saturation of the magnetization, 
namely $M_{\rm s}=3L/2$ for this system. 
$E(L,M)$ is calculated by the Lanczos algorithm under the 
periodic boundary condition ($ {\bf S}_{N+1}={\bf S}_1$) 
and the twisted boundary condition 
($S^{x,y}_{N+1}=-S^{x.y}_1, S^z_{N+1}=S^z_1$), 
for $L=$4, 6 and 8. 
Under the twisted boundary condition 
we calculate the lowest energy $E_{{\rm TBC}P=+}(L,M)$
($E_{{\rm TBC}P=-}(L,M)$) 
in the subspace where 
the parity is even (odd) with respect to the lattice inversion 
at the twisted bond.

\section{Magnetization plateau}

In the isotropic coupling case $\lambda=1$ 
the previous theoretical study indicates that 
the model (2) exhibits the 1/3 magnetization plateau 
due to two different mechanisms depending on the 
parameters $J_1$, $J_2$, $J_3$ \cite{okamoto2}. 
Particularly $J_2$ plays a key role. 
The plateau is based on the ferrimagnet-like mechanism for 
smaller $J_2$, while it is based on the structure where 
the singlet dimer lies on the $J_2$ bond and the other spins are free 
for larger $J_2$. 
The small $J_2$ plateau and the large $J_2$ one 
are denoted as the plateaux A and B, respectively. 
The second-order quantum phase transition occurs 
on the boundary between the two plateau phases. 
In this paper the quantum phase transition 
from the plateau phase to the no-plateau one 
with respect to the $XY$-like coupling anisotropy 
is investigated in the both cases of the plateaux A and B. 

\section{Level spectroscopy analysis and phase diagram}

\begin{figure}[tbh]
\centerline{\includegraphics[scale=0.4]{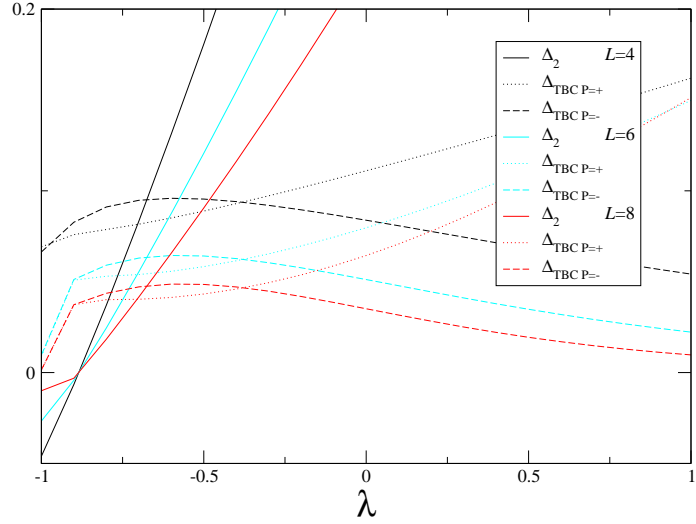}}
\caption{$\lambda$ dependence of the three gaps 
$\Delta _2$(solid line), $\Delta_{{\rm TBC}P=+}$(dotted line) 
and $\Delta_{{\rm TBC}P=1}$(dashed line) 
for $J_1=1.0$, $J_2=1.1$ and $J_3=0.1$. 
Black, blue and red lines correspond to the system sizes 
$L=$4, 6 and 8, respectively. 
}
\label{LS}
\end{figure}

In order to detect the quantum phase transitions among the 
plateaux-A, -B and no-plateau phases, 
the level spectroscopy analysis\cite{kitazawa1,kitazawa2} is one of the best methods. 
According to this analysis, 
we should compare the following three energy gaps; 
\begin{eqnarray}
\label{delta2}
&&\Delta _2 ={E(L,M-2)+E(L,M+2)-2E(L,M) \over 2}, \\
\label{tbc+}
&&\Delta_{{\rm TBC}P=+}=E_{{\rm TBC}P=+}(L,M)-E(L,M), \\
\label{tbc-}
&&\Delta_{{\rm TBC}P=-}=E_{{\rm TBC}P=-}(L,M)-E(L,M).
\end{eqnarray}
The level spectroscopy method indicates that the smallest gap 
among these three gaps for $M=L=M_{\rm s}/3$ determines the phase 
at $m=1/3$. 
$\Delta_2$, $\Delta_{{\rm TBC}P=+}$ and $\Delta _{{\rm TBC}P=-}$ 
correspond to the no-plateau, plateau-A and plateau-B phases, 
respectively. 
Especially, $\Delta_{{\rm TBC}P=+}$ and  $\Delta_{{\rm TBC}P=-}$ directly reflect the symmetry of two plateau states.
The $\lambda$ dependence of these three gaps for $J_1=1.0$, $J_2=1.1$ 
and $J_3=0.1$ is shown in Fig. \ref{LS} for $L=$4, 6 and 8. 

\bigskip
\bigskip

\begin{figure}[tbh]
\centerline{\includegraphics[scale=0.4]{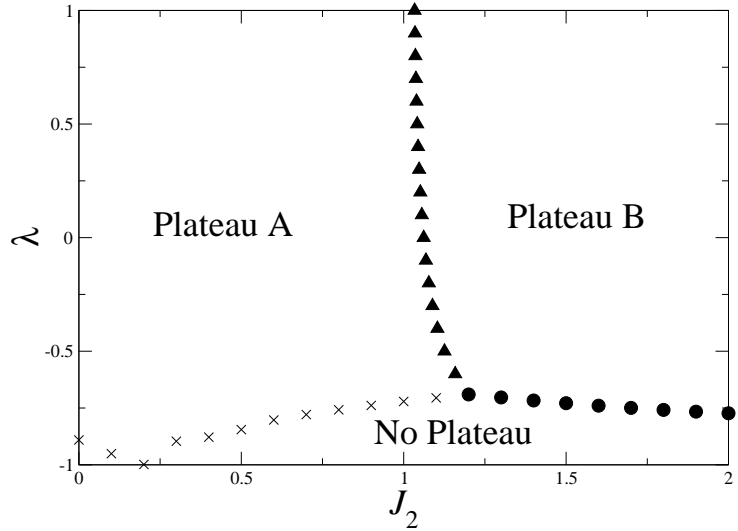}}
\caption{
Phase diagram at $m=1/3$ for $J_1=1.0$ and $J_3=0.1$. 
Crosses, circles and triangles correspond to the boundaries 
between the plateau-A and no-plateau, between the plateau-B and no-plateau, 
and between the plateaux-A and -B phases, respectively. 
}
\label{phase}
\end{figure}

Assuming that the finite-size correction is proportional to $1/L^2$, 
we estimate the phase boundaries in the thermodynamic limit 
from every level-cross point. 
The phase diagram in the $J_2$-$\lambda$ plane at $m=1/3$ 
for $J_1=1.0$ and $J_3=0.1$ 
is shown in Fig. \ref{phase}.

It suggests that both plateaux are so stable against the $XY$-like 
anisotropy that the plateaux vanish at quite large negative $\lambda$. 

\section{Concluding Remarks}

Using the numerical exact diagonalization and the level spectroscopy 
analysis, the $S=1/2$ distorted diamond spin chain is investigated. 
It is found that the 1/3 magnetization plateau vanishes for sufficiently 
large negative $XY$-like anisotropy, in both cases of the ferrimagnet-like 
and the dimer-monomer plateaux. 
A typical phase diagram at $m=1/3$ is presented in  Fig.\ref{phase}. 

One of the remarkable natures of the plateau phase diagram Fig.\ref{phase} is the survival of the
1/3 plateau to the $\lambda <0$ region,
where the direct $S^z-S^z$ interaction is ferromagnetic. 
The essential mechanism of the plateau-A is the Lieb-Mattis ferrimagnetism\cite{LM}. 
In the simple $S=1/2$ $XXZ$ chain with $\lambda <0$,
although the direct $S^z-S^z$ interaction is ferromagnetic,
the antiferromagnetic $S^z-S^z$ correlation survives to $\lambda = -1$
induced by the antiferromagnetic $S^x-S^x$ and $S^y-S^y$ correlations.
Thus, the plateau A survives to near $\lambda = -1$,
where the ferromagnetic long-range order begins.
On the other hand, the plateau-B state is attributed to the singlet-dimer formation at the $J_2$ bonds,
which makes the nodal spins nearly free.
This type of singlet dimer survives to $\lambda = -1/\sqrt{2} = -0.707$
in the $S=1/2$ bond-alternating $XXZ$ chain\cite{kohmoto,oka-sugi}.
Further, if we neglect $J_3$, the present system becomes a trimerized $S=1/2$ $XXZ$ chain,
in which Okamoto and Kitazawa\cite{oka-kita} showed that the no-plateau state appears
for $\lambda < -0.729$.
These facts well explains why the boundary of the platea-B state and no-plateau state
lies around $\lambda = 0.70 \sim 0.75$.
Thus both plateaus survive into the $\lambda <0$ region,
and the plateau-A is more stable than the plateau-B as the $\lambda = -1$ line is approached.

We think that the no-plateau state is essentially the Tomonaga-Luttinger liquid state in the magnetic field.
We can see a sudden decrease of the phase boundary between the plateau-A and the no-plateau $J_2=0.2$.
The physical explanation for this phenomena is a future problem.
We believe that our plateau phase diagram is important for the full understanding of the distorted diamond chain
and will yields important informations if the distorted diamond chain with the $XXZ$ anisotropy
is found or synthesized in future.

\section*{Acknowledgment}
This work was partly supported by JSPS KAKENHI, Grant Numbers 16K05419, 
16H01080 (J-Physics) and 18H04330 (J-Physics). 
A part of the computations was performed using 
facilities of the Supercomputer Center, 
Institute for Solid State Physics, University of Tokyo, 
and the Computer Room, Yukawa Institute for Theoretical Physics, 
Kyoto University.

\end{document}